\documentclass[aps, prl, twocolumn, superscriptaddress]{revtex4}
\usepackage{graphicx}
\usepackage{array}
\usepackage{amsmath,amssymb}
\usepackage{mathrsfs}
\usepackage[dvipsnames,usenames]{color}
\usepackage{float}
\newcommand{\bm}[1]{\mbox{\boldmath{$#1$}}}
\begin{document}
\title{Frustration-enhanced persistent currents in correlated trimer nanorings}
\author{Tie-Feng Fang}
\email[]{fangtiefeng@lzu.edu.cn}
\affiliation{School of Physics and Technology, Nantong University, Nantong 226019, China}
\author{Wei-Tao Lu}
\affiliation{School of Physics and Technology, Nantong University, Nantong 226019, China}
\author{Ai-Min Guo}
\affiliation{Hunan Key Laboratory for Super-microstructure and Ultrafast Process, School of Physics and Electronics,
Central South University, Changsha 410083, China}
\author{Qing-Feng Sun}
\email[]{sunqf@pku.edu.cn}
\affiliation{International Center for Quantum Materials, School of Physics, Peking University, Beijing 100871, China}
\affiliation{Hefei National Laboratory, Hefei 230088, China}
\date{\today}
\begin{abstract}
We investigate the persistent current in a correlated trimer nanoring comprising of three magnetic atoms, which sits on a metallic host and encloses a magnetic flux. In the molecular-orbital regime, charge fluctuations can reverse the aromaticity of the trimer molecule by driving quantum phase transitions between many-body states. It is shown that in frustrated trimers the superexchange-induced current as a function of interatom hopping is enhanced by the competition of conflicting magnetic orders, with an asymmetric peak at the quantum criticality separating the ferromagnetic and antiferromagnetic Kondo regimes. Interestingly, the critical current undergoes an anomalous rise with temperature before decaying, signaling the suppression of Kondo bound state at finite temperature. Our results demonstrate that the coherent current response to external flux indeed conveys important information on the states of strongly correlated systems.
\end{abstract}
\maketitle

One of the most striking manifestations of quantum coherence in systems with broken time-reversal symmetry is the emergence of equilibrium currents \cite{Buttiker1983, Imry1999, Imry2002} flowing dissipationlessly through annular geometries due to the wave-function phase winding around the loop. Besides accessing quantum phases of matter, such persistent currents (PCs) have many applications from precision sensing to quantum computing. They are ubiquitous phenomena observed in different fields including condensed-matter physics \cite{Levy1990}, aromatic chemistry \cite{Gomes2001}, ultracold atomic gases \cite{Ryu2007}, and microcavity polaritonics \cite{Lukoshkin2018}.

Early studies of PCs were associated with quantized flux in superconducting cylinders \cite{Deaver1961, Byers1961, Onsager1961, Doll1961}. Intense research efforts later focused on small metallic rings pierced by magnetic fields, where the PC was measured as the hallmark of mesoscopic physics \cite{Saminadayar2007, Kleemans2007, Bluhm2009, BJ2009, CB2013}. Recent advances in atomtronic matter-wave circuits and characterizing aromatic molecules have rekindled interest in the field with new scope. By virtue of the enhanced flexibility with tunable interactions, atomic PCs have been generated \cite{Wright2013, Beattie2013, Cai2022, Pace2022, Amico2022, Tononi2023, Simjanovski2023} in ultracold bosonic and fermionic gases throughout the BCS to BEC crossover, which highlights the coherent phase profiles of many-body collective circulation states. PC is also a defining notion for the aromaticity of cyclic molecules. The H\"{u}ckel's rule \cite{Gomes2001, Huckel1931, Sola2022} asserts that molecular nanorings with an odd (even) number of spin-up and of spin-down electrons are aromatic (antiaromatic), sustaining diatropic (paratropic) PCs when placed in magnetic fields. This empirical rule takes account of shell-filling effects based on a simple molecular orbital theory and in fact coincides with the magnetic responses of noninteracting electronic rings \cite{Cheung1988, Loss1991}. While the rule was recently shown to extend to much larger macrocycles \cite{Peeks2017, Rickhaus2020, Parmar2022, Rahav2023, Ren2023}, there are situations \cite{Sola2022, Reig2021, Saha2023, Merino2023} that do not fall into this category and hence indicate the important influences of interaction and decoherence on molecular ring currents.

On the other hand, theoretical insights on PCs in strongly correlated systems are insufficient \cite{Zvyagin1995}. Previous theories revealed the distinct scalings of PCs in a noninteracting ring with a Kondo impurity \cite{Sorensen2005}, while the PC in a strongly interacting ring was enhanced due to the chirality of a magnetic impurity \cite{Schlottmann2003}. A sign change in PCs, corresponding to a reversal of aromaticity, was recently demonstrated by multireference calculations of electron-vibration interactions in the cyclocarbon $\mathrm{C_{16}}$ \cite{Roncevic2023}. It was predicted that the circular Hubbard model can sustain PCs exhibiting the temperature-dependent periodicity \cite{Patu2022}. However, interesting issues like the critical behavior of PCs near quantum phase transitions (QPTs), detailed information on many-body states conveyed by PCs at the single quantum level, and the environmental effect have not yet been fully analyzed.

In this Letter, we address these issues by studying PCs in a magnetic-impurity trimer nanoring threaded by a magnetic field. The trimer is apically coupled to a metallic host that renders the system strongly correlated but also serves as a decoherent environment. By using the numerical renormalization group (NRG) method, we reveal fascinating current characteristics arising from the competition of magnetic ordering, geometric frustration, and Kondo correlations. While the aromaticity of the trimer molecule may be reversed by QPTs, higher-order currents in the frustrated trimer mediated by superexchange processes are always paratropic. In particular, the superexchange current is enhanced and reaches a maxima at the frustrating quantum criticality that separates the ferromagnetic and antiferromagnetic Kondo regimes. Being vulnerable to decoherence, the critical current maxima exhibits anomalous temperature dependence due to the Kondo fading away. These features indicate that the PC is a sensitive probe for correlated quantum phases of many-body systems.

Specifically, our trimer nanoring is modeled by the Hamiltonian $H=H_0+H_\mathrm{env}$,
\begin{eqnarray}
&&\hspace{-0.7cm}H_{0}=\sum_{j=1,2,3}\sum_{\sigma=\uparrow,\downarrow}\Big(\varepsilon_0
d^\dagger_{j\sigma}d_{j\sigma}
+\frac{1}{2}Ud^\dagger_{j\sigma}d_{j\sigma}d^\dagger_{j\bar{\sigma}} d_{j\bar{\sigma}}\Big)\nonumber\\
&&\ -\sum_\sigma\big(te^{-i\varphi}d^\dagger_{2\sigma}d_{1\sigma}
+t'd^\dagger_{3\sigma}d_{2\sigma}+td^\dagger_{1\sigma}d_{3\sigma}+\mathrm{H.c.}\big),\\
&&\hspace{-0.75cm}H_\mathrm{env}=\sum_{k,\sigma}\varepsilon_kC^\dagger_{k\sigma}C_{k\sigma}
+\sum_{k,\sigma}V(C^\dagger_{k\sigma}d_{1\sigma}+\mathrm{H.c.}).
\end{eqnarray}
The three-site circular Hubbard term $H_0$ represents the isolated trimer ring, where $d^\dagger_{j\sigma}$ creates an electron of energy $\varepsilon_0$ and spin $\sigma$ in the $j$th magnetic atoms, $U$ denotes the onsite interaction, $t$ and $t'$ are the interatom hoppings, and a magnetic flux $\Phi$ induces the phase $\varphi=2\pi\Phi/\Phi_0$ with $\Phi_0=h/e$ the flux quantum. $H_\mathrm{env}$ describes electrons in the metallic environment and their coupling $V$ to site $1$ which is parameterized to $\Gamma\equiv\pi\rho V^2$ with $\rho$ the density of host states. At $\varphi=0$, rich phase diagrams of this model have already been elucidated \cite{Mitchell2009, Mitchell2013, Wojcik2020}. Nevertheless, the coherent current response to the magnetic flux remains unexplored. Note that the presence of magnetic flux breaks the parity symmetry that $H$ is invariant under a permutation of sites $2$ and $3$. On the other hand, specific values of magnetic flux, i.e., $\varphi=(n+\frac{1}{2})\pi$ ($n\in\mathbb{Z}$), can restore the particle-hole symmetry that is broken at $\varphi=0$. Intriguing current characteristics are thus expected due to the dual role of the magnetic flux as both symmetry breaking and restoring perturbations.

To analyze the trimer ring current subject to the environmental effect, we have solved the full Hamiltonian $H$ by using the full density-matrix NRG method \cite{Wilson1975, Krishna-murthy1980, Bulla2008, Anders2005, Peters2006, Weichselbaum2007, Fang2015}, while the isolated trimer $H_0$ can be readily sovled using exact diagonalization. We are interested in the following observables: i) the PC in the trimer nanoring, $I$, calculated as
\begin{equation}
I=\frac{ie}{\hbar}\sum_\sigma\Big(te^{-i\varphi}\big\langle d^\dagger_{2\sigma}d_{1\sigma}\big\rangle-te^{i\varphi}\big\langle d^\dagger_{1\sigma}d_{2\sigma}\big\rangle\Big);
\end{equation}
ii) the spectral density on the $1$st impurity
$A(\varepsilon)=-\frac{1}{\pi}\mathrm{Im}\langle\langle d_{1\sigma};d^\dagger_{1\sigma}\rangle\rangle$; iii) the spin correlation $\langle\bm{S}_2\cdot\bm{S}_3\rangle$ between the 2nd and 3rd impurities; and iv) the total number of electrons in the trimer, $N=\sum_{j,\sigma}\langle d^\dagger_{j\sigma}d_{j\sigma}\rangle$. NRG algorithms were implemented using a discretization parameter $\Lambda=10$ for $I$, $\langle\bm{S}_2\cdot\bm{S}_3\rangle$, and $N$, and $\Lambda=5$ for $A(\varepsilon)$, and retaining $M_K=1600\sim2400$ states per iteration. We take the half bandwidth $D=1$, the Fermi energy $E_F=0$, the temperature $T=0$ (unless stated otherwise), and the coupling of site $1$ with the other two sites fixed to $t=0.01$. We also assume $2\varepsilon_0+U=0$ such that the trimer PC has the period of half flux quantum, $I(\varphi)=I(\varphi+\pi)$, and vanishes $I(\frac{n}{2}\pi)=0$. It is thus sufficient to discuss the PC only within the flux range of $0<\varphi<\frac{\pi}{2}$. The trimer molecule is aromatic or antiaromatic if its PC in this range is diatropic ($I<0$) or paratropic ($I>0$), according to the magnetic criterion for aromaticity \cite{GP2015}.

\begin{figure}
\hspace{2.0mm}
\includegraphics[width=1.08\columnwidth]{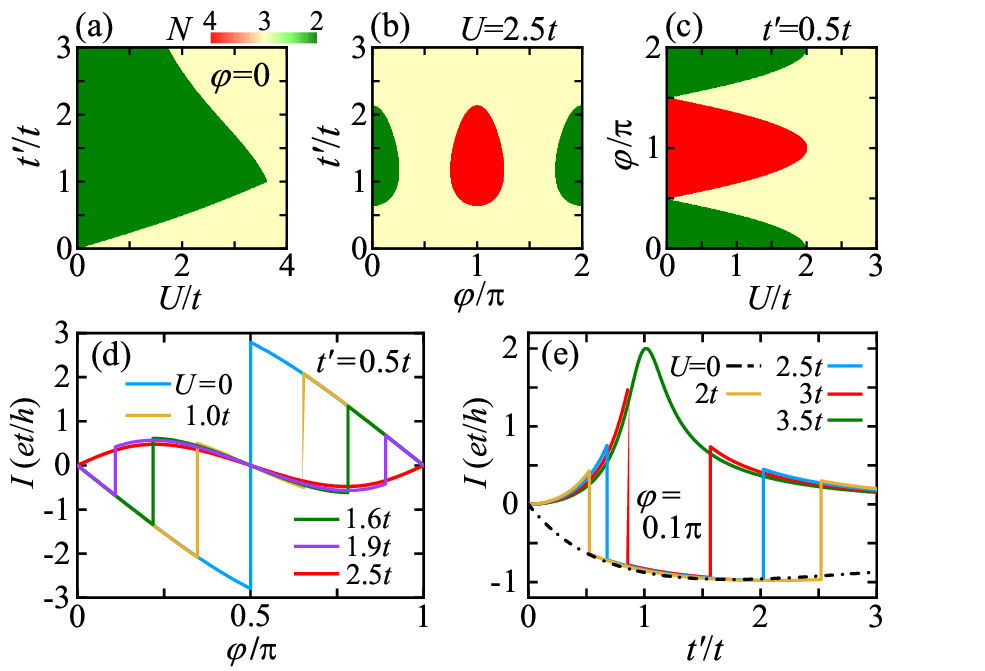}
\caption{Results for isolated trimer molecules with $\Gamma=0$. (a)-(c) Number $N$ of electrons occupying the trimer as a function of the hopping $t'$, the interaction $U$, and the flux phase $\varphi$. (d) Phase-dependent PCs, $I(\varphi)$, for different $U$ at $t'=0.5t$. (e) PCs, $I(t')$, as a function of $t'$ for different $U$ at $\varphi=0.1\pi$.}
\end{figure}

We first consider the molecular-orbital regime occurring for weak or not so strong $U$. In the absence of metallic environment, the trimer occupation $N$ varies abruptly among $2$, $3$, and $4$, depending on the interaction $U$, the hopping $t'$ between sites $2$ and $3$, and the flux phase $\varphi$, as shown in Figs.\,1(a)-1(c). Note that the trimer geometry intrinsically breaks the particle-hole symmetry. A half filling occurs only for $\varphi$ approaching $\frac{\pi}{2}$ or $\frac{3}{2}\pi$, strong $U$, or large hopping anisotropy ($t'/t\ll1$ or $\gg1$), because the particle-hole symmetry is strictly restored in the limits of $\varphi=(n+\frac{1}{2})\pi$, $U\to\infty$, and $t'/t\to 0$ or $\infty$. Figure 1(d) displays the phase-dependent PCs for different interactions at anisotropic hoppings. In the $U=0$ case, the PC exhibits the typical sawtooth behavior [the blue line in Fig.\,1(d)] \cite{Cheung1988, Loss1991}. The trimer molecule is aromatic and $1/3$-filled with $N_\uparrow=N_\downarrow=1$ for $0<\varphi<\frac{\pi}{2}$, thereby obeying the H\"{u}ckel's rule \cite{Sola2022, Merino2023}. For large $U$, the PC becomes sinusoid-like and the half-filled trimer is antiaromatic [the red line in Fig.\,1(d)]. But the H\"{u}ckel's rule is not applicable to this $N=3$ case. We emphasize that at $\Gamma=0$, the $U$-driven transition from sawtooth to sinusoidal, as shown in Fig.\,1(d), only occurs for $t'\ne t$. The Bethe ansatz solution \cite{Zvyagin1995, Zvyagin1990} for a Hubbard ring with equal hoppings indicated a sawtooth PC for arbitrary $U$. Interestingly, sudden reversals of aromaticity occur at some $\varphi$ values within the first quadrant for intermediate $U$ [the yellow, green, and purple lines in Fig.\,1(d)], corresponding to a jump from $N=2$ to $3$. Such aromaticity reversals also appear in the PC as a function of the hopping $t'$ [Fig.\,1(e)]. For finite $U$, large hopping anisotropy favors the half-filling state, giving rise to paratropic trimer currents. Upon decreasing the anisotropy by tuning the hoppings to $t'/t\sim1$ from both the $t'<t$ and $t'>t$ sides, the trimer undergoes steep transitions to the $1/3$-filling state carrying aromatic ring currents. For sufficiently large $U$, the diatropic PC valley in $I$ vs $t'$ disappears, resulting in a prominent paratropic current peaked at $t'/t\sim1$ [Fig.\,1(e)].

\begin{figure}
\hspace{-3mm}
\includegraphics[width=1.0\columnwidth]{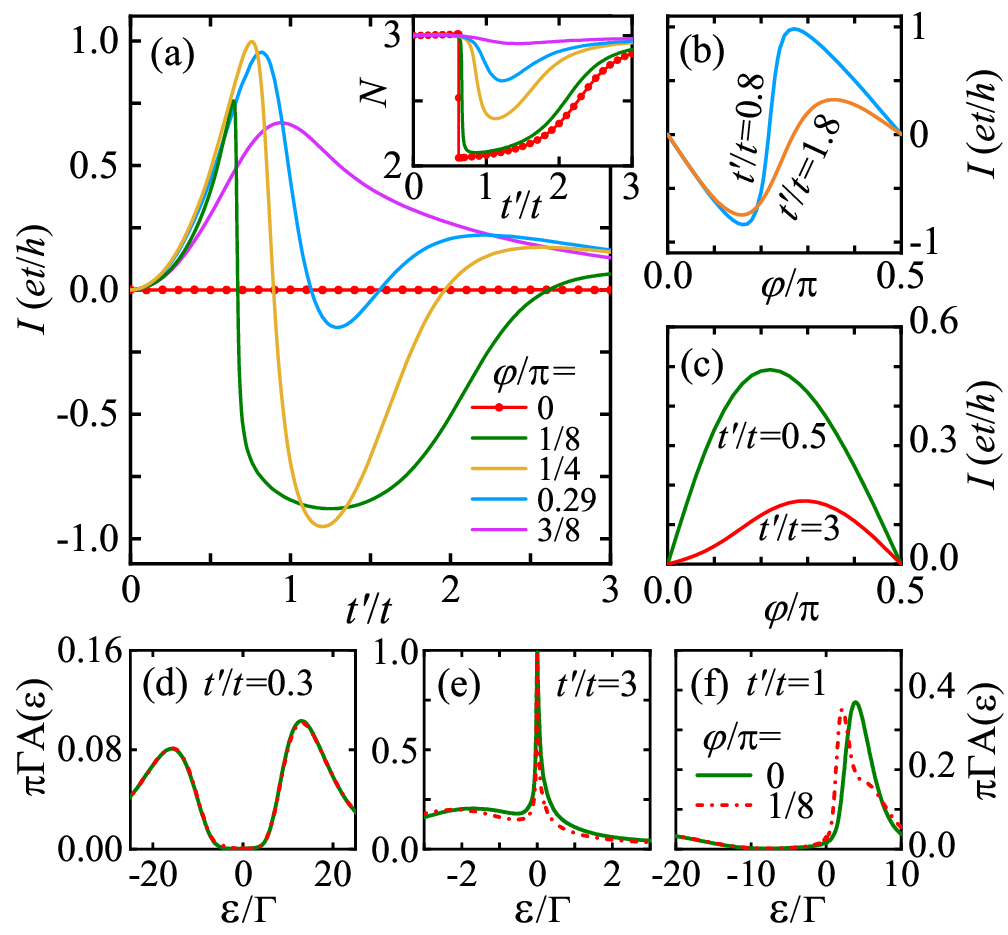}
\caption{Aromaticity reversal of trimer molecules coupled to metallic environment. (a) Evolution of the current $I$ vs the hopping $t'$ with variation of the flux phase $\varphi$. Inset: Corresponding evolution of the electron number $N$. (b) and (c) Phase-dependent PCs, $I(\varphi)$, for various hoppings $t'$. (d)-(f) Spectral density $A(\varepsilon)$ on site $1$ with several values of the hopping $t'$ and phase $\varphi$. Results are for $U=2.5t$ and $\Gamma=0.1t$.}
\end{figure}

Figure 2 elaborates on how the metallic environment affects the aromaticity reversal of the trimer molecule. It is shown that even for small flux $\varphi$, the trimer-metal coupling has already smoothen the two steep aromaticity reversals on the $t'<t$ and $t'>t$ sides, respectively [Fig.\,2(a)]. Such aromaticity reversals can also be driven by the flux, provided that proper hoppings are chosen [Fig.\,2(b)]. Otherwise, the trimer keeps antiaromatic [Fig.\,2(c)]. We emphasize that in the presence of environment, the underlying physics of the aromaticity reversals on the $t'<t$ and $t'>t$ sides are distinct. The low-lying states of the half-filled trimer achieved in large hopping anisotropy [inset of Fig.\,2(a)] constitute a spin doublet. For $t'\ll t$, the effective model describing the doublet trimer with the environment is the ferromagnetic Kondo model \cite{Mitchell2009, Mitchell2013} whose ground state is a free local moment \cite{ Koller2005, Mehta2005} characterized by the logarithmically vanishing spectral density $A(E_F)=0$ on site $1$ [Fig.\,2(d)]. However, the antiferromagnetic Kondo effect arising for $t'\gg t$ drives the system to the screened singlet ground state \cite{Mitchell2009, Mitchell2013}, causing a sharp Kondo resonance in the spectral density [Fig.\,2(e)]. In the $1/3$-filling valley realized for $t'\sim t$ and small $\varphi$ [inset of Fig.\,2(a)], the singlet trimer state effectively decouples from the environment at zero temperature, indicated by a finite but featureless spectral density near the Fermi energy [Fig.\,2(f)]. The resulting ground state is continuously connected to the Kondo singlet at $t'\gg t$ but not to the doublet at $t'\ll t$. Therefore, the aromaticity reversal of PCs on the $t'<t$ side is due to a singlet-doublet QPT, while it is essentially a singlet-singlet crossover on the $t'>t$ side. The QPT is of Kosterlitz-Thouless type due to the symmetry-breaking effect caused by the magnetic flux $\varphi$ (we will demonstrate this in a different scenario later). For $\varphi=0$ and thus $I=0$, however, the parity symmetry protects the QPT being a first-order level crossing [the red curve in inset of Fig.\,2(a)] \cite{Mitchell2013, Wojcik2020}.

Remarkably, as the flux $\varphi$ increases to approach $\frac{\pi}{2}$, the restoration of the particle-hole symmetry elevates the $1/3$-filling valley in $N$ vs $t'$ [inset of Fig.\,2(a)]. Meanwhile, the diatropic PC valley in $I$ vs $t'$ dramatically transforms into a paratropic peak around $t'/t\sim1$ [Fig.\,2(a)]. The peak is noticeably asymmetric, similar to the one driven by the interaction as shown in Fig.\,1(e).

To glean analytical insights into the nature of this prominent current peak, we take a sufficiently strong interaction $U$ such that the trimer is always half-filled with each site being singly occupied. In this subspace, by taking account of virtual charge excitations up to third order in $t$, $t'$, and $V$, we obtain the effective spin model \cite{Fradkin2013, Wang2020, Bulaevskii2008, Scarola2004, SM}
\begin{equation}
\mathscr{H}=J_\mathrm{K}\bm{S}_1\cdot\bm{S}+J\bm{S}_1\cdot(\bm{S}_2
+\bm{S}_3)+J'\bm{S}_2\cdot\bm{S}_3
+\lambda\bm{\Delta}_{123}.
\end{equation}
Here $\bm{S}_j$ is the spin on site $j$, $\bm{\Delta}_{123}=\bm{S}_1\cdot(\bm{S}_2\times\bm{S}_3)$, and $\bm{S}$ the spin of environmental electrons, with $J=4t^2/U$, $J'=4t'^2/U$, $\lambda=(24t^2t'/U^2)\sin\varphi$, and $J_\mathrm{K}=4\Gamma/(\pi\rho U)$ defining the Kondo temperature $T_\mathrm{K}=\sqrt{(\Gamma U/2)}e^{-1/(2\rho J_\mathrm{K})}$. In this frustrated-trimer regime, PCs along the trimer loop arise from the third-order hopping processes due to the superexchange interaction $\lambda\bm{\Delta}_{123}$ mediated by virtual charge excitations \cite{note}. The ground state of the isolated trimer up to second order, $\mathscr{H}_{\text{2nd}}=J\bm{S}_1\cdot(\bm{S}_2+\bm{S}_3)+J'\bm{S}_2\cdot\bm{S}_3$, is an even-parity doublet $|+;\sigma\rangle\equiv(1/\sqrt{2})(|\sigma\hspace{-0.8mm}\uparrow\downarrow\rangle -|\sigma\hspace{-0.8mm}\downarrow\uparrow\rangle)$ for $t'>t'_c$ or an odd-parity one $|-;\sigma\rangle\equiv(\sigma/\sqrt{6})(|\sigma\hspace{-0.8mm}\uparrow\downarrow\rangle +\left|\sigma\hspace{-0.8mm}\downarrow\uparrow\right\rangle-2|\bar\sigma\sigma\sigma\rangle)$ for $t'<t'_c$ \cite{Mitchell2009, Mitchell2013}. Their energies $E_+=-\frac{3}{4}J'$ and $E_-=\frac{1}{4}J'-J$ cross at the critical hopping $t'_c=t$. In $|\pm;\sigma\rangle$, sites $2$ and $3$ form a local singlet (triplet). The parity symmetry-breaking perturbation $\lambda\bm{\Delta}_{123}$ of the magnetic flux can then mix these magnetic orders, leading to an avoided level crossing. In detail, we obtain the doublet ground state $|g;\sigma\rangle$ and its energy $E_g$ of the isolated trimer up to third order, $\mathscr{H}_{\text{3rd}}=\mathscr{H}_\text{2nd}+\lambda\bm{\Delta}_{123}$, as $|g;\sigma\rangle=\alpha|+;\sigma\rangle+\beta|-;\sigma\rangle$ and $E_g=E_--\frac{1}{2}\delta/(J'-J)$ \cite{SM}, where $|\alpha|^2=1-|\beta|^2=\frac{8\delta+3\lambda^2}{8\delta+6\lambda^2}$ and $\delta=(J'-J)^2+(J'-J)\sqrt{(J'-J)^2+3\lambda^2/4}$. The effective Kondo model for $T_\text{K}\ll J$ can then be derived as $J_\text{K}\sum_{\sigma,\,\sigma'}|g;\sigma\rangle\langle g;\sigma|\bm{S}_1\cdot\bm{S}|g;\sigma'\rangle\langle g;\sigma'|=\widetilde{J}_\text{K}\widetilde{\bm{S}}_1\cdot\bm{S}$, with $\widetilde{J}_\text{K}=(\frac{4}{3}|\alpha|^2-\frac{1}{3})J_\text{K}$ and $\widetilde{\bm{S}}_1$ the spin operator for the doublet $|g;\sigma\rangle$. At zero magnetic flux, for $t'>t'_c$ and thus $|\alpha|^2=1$, the interaction $\widetilde{J}_\text{K}=J_\text{K}$ is antiferromagnetic, while $\widetilde{J}_\text{K}=-\frac{1}{3}J_\text{K}$ is ferromagnetic because of $|\alpha|^2=0$ for $t'<t'_c$ \cite{Mitchell2009, Mitchell2013}. The discontinuity in this level-crossing QPT will be removed by the magnetic flux. For $\varphi\ne0$, as the hopping $t'$ decreases across $t'_c$, the effective coupling $\widetilde{J}_\text{K}$ continuously diminishes from $J_\text{K}$ to $0$ and then to $-\frac{1}{3}J_\text{K}$, in accordance with $|\alpha|^2$ continuously reduced from $1$ to $0$ \cite{SM}. Such a vanishing of the Kondo scale $\widetilde T_\text{K}\sim e^{-1/\rho\widetilde{J}_\text{K}}$ as $\widetilde J_\text{K}\to0$ is characteristic of the Kosterlitz-Thouless QPT \cite{Kosterlitz1973}. The most striking feature of PCs in the above physical scenario already shows up in the isolated trimer,
\begin{equation}
I=-\frac{\partial E_g}{\partial \Phi}=\frac{3\pi e}{4h}\frac{\lambda^2\cot\varphi}{\sqrt{(J'-J)^2+3\lambda^2/4}},
\end{equation}
indicating a current peak at $t'=t'_c=t$ in $I$ vs $t'$ where $E_g$ acquires a maximal $\Phi$-dependence \cite{SM}. The peak is asymmetric due to the dependence of $\lambda$ on $t'$.

\begin{figure}
\hspace{0mm}
\includegraphics[width=1.0\columnwidth]{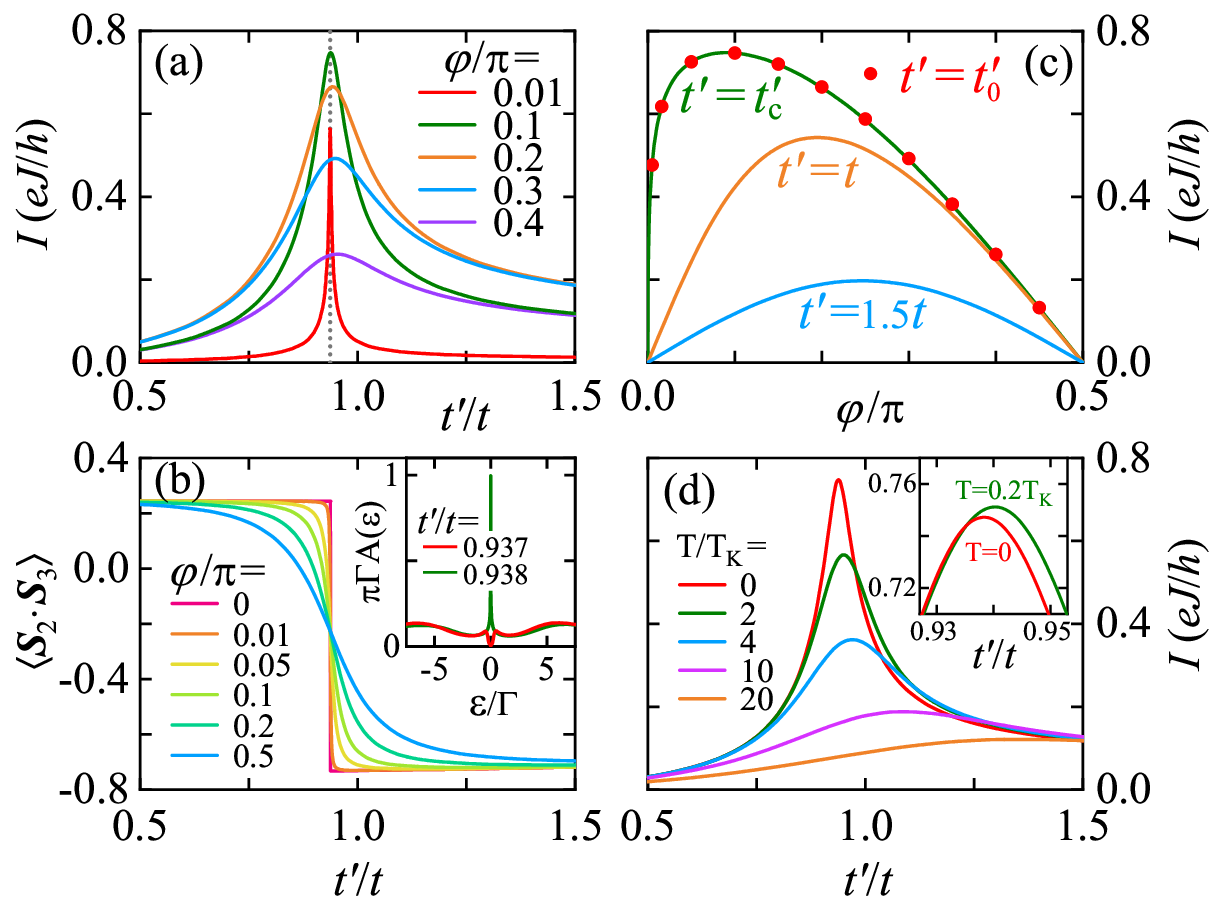}
\caption{(a) PC $I$ and (b) spin correlation $\langle\bm{S}_2\cdot\bm{S}_3\rangle$ as functions of the hopping $t'$ with variation of the flux phase $\varphi$. For $\varphi=0$, the spin correlation abruptly changes at $t'=t'_c\equiv0.937442t$. The critical $t'_c$ is marked in (a) by the dotted line. Inset of (b): Zero-field spectral density for $t'$ close to $t'_c$. (c) Phase-dependent PCs for different $t'$, with $t'_0$ being the peak position in $I$ vs $t'$. (d) and inset: Temperature evolution of the current peak in $I$ vs $t'$ at $\varphi=0.1\pi$. Results are for frustrated trimers with $U=16t$ and $\Gamma=t$ from which $T_K=0.00528t$ and $J=0.25t$.}
\end{figure}

These analytical insights are verified by our full NRG results of frustrated trimers for very strong $U$ \cite{note} and $T_\mathrm{K}\ll J$, as shown in Fig.\,3. Unlike the molecular-orbital case, the paratropic current peak in $I$ vs $t'$ with typical asymmetric lineshape, is now robust to the flux, which becomes sharper as $\varphi$ decreases [Fig.\,3(a)], consistent with the peak width determined by $\lambda\propto\sin\varphi$ predicted from Eq.\,(5). Indeed, the current peak is located at the critical hopping $t'_c$ indicated by the discontinuity in the $\varphi=0$ spin correlation, though $t'_c$ is now out of $t$ due to the environment effect. For $t'<t'_c$ ($t'>t'_c$), $\bm{S}_2$ and $\bm{S}_3$ form a local triplet (singlet) with $\langle\bm{S}_2\cdot\bm{S}_3\rangle\simeq\tfrac{1}{4}$ ($-\frac{3}{4}$) [Fig.\,3(b)]. At $t'=t'_c$, the frustration of these two magnetic orders results in a level-crossing QPT separating the ferromagnetic Kondo regime characterized by $A(E_F)=0$ from the antiferromagnetic Kondo state with a sharp resonance in the spectral density [inset of Fig.\,3(b)] \cite{Mitchell2009, Mitchell2013}. The magnetic flux, like other symmetry-breaking perturbations \cite{Mitchell2009, Mitchell2013, Baruselli2013}, turns the QPT into a Kosterlitz-Thouless transition, as characterized by the broadened step in the spin correlation [Fig.\,3(b)]. The resulting frustrated ground state of the trimer system can then sustain a large loop current near the criticality.

Notably, the height of the PC peak in $I$ vs $t'$ matches well with the critical current at $t'_c$ in the full phase range and the phase-dependent current evolves from sawtooth-like to sinusoid-like as the hopping $t'$ deviates from $t'_c$ [Fig.\,3(c)]. These distinct characteristics demonstrate that the frustration-enhanced PC peak is indeed a good probe to the QPT of the trimer. Interestingly, while the current peak easily decays upon increasing temperature [Fig.\,3(d)], there is a small interval of temperature where the peak increases counterintuitively [inset of Fig.\,3(d)].

\begin{figure}
\hspace{0mm}
\includegraphics[width=1.0\columnwidth]{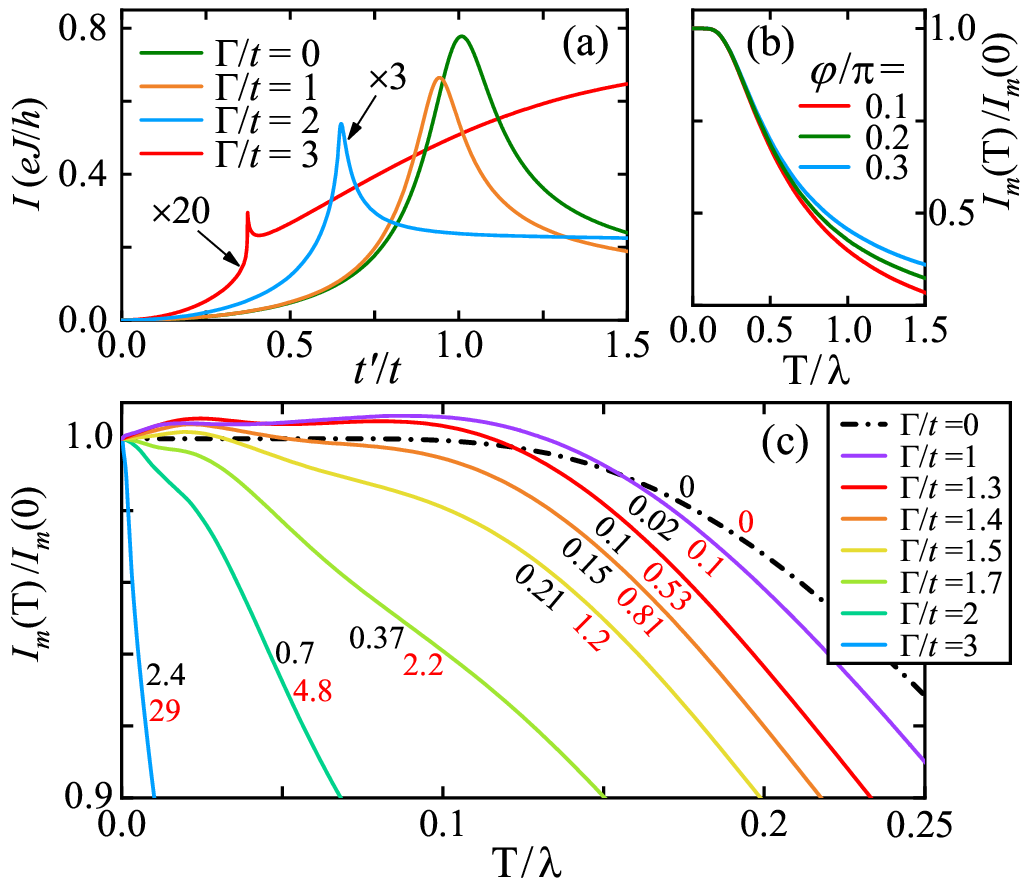}
\caption{(a) Evolution of the current $I$ as a function of the hopping $t'$ by varying the trimer-environment coupling $\Gamma$, at the phase $\varphi=0.2\pi$. (b),(c) Temperature dependence of the current-peak height $I_m(T)$ in $I$ vs $t'$, for (b) different $\varphi$ at $\Gamma=0$ and (c) different $\Gamma$ at $\varphi=0.2\pi$. Data are extracted by tracing the $T$-dependent peak positions, while the three-order superexchange $\lambda$, used to scale $T$, is determined by taking $t'$ at the zero-$T$ peak position. In (c), the black and red numbers next to each curve indicate the values of $T_\mathrm{K}/J$ and $T_\mathrm{K}/\lambda$, respectively. Results are for frustrated trimers with $U=16t$ and $J=0.25t$.}
\end{figure}

To explore further this anomalous increase, we plot in Fig.\,4 the evolution of current peak with the trimer-environment coupling $\Gamma$. Note first that as $\Gamma$ increases, the zero-$T$ current peak gets strongly suppressed [Fig.\,4(a)] and the finite-$T$ peak decays faster [Fig.\,4(c)], manifesting the environment-induced decoherence effect. There are also a leftward shift and an evident sharpening of the current peak as seen in Fig.\,4(a). This is because the indirect ferromagnetic interaction between $\bm{S}_2$ and $\bm{S}_3$ decreases with $\Gamma$, thereby reducing the direct antiferromagnetic exchange $J'$ needed to reach the frustrating critical point. Meanwhile, the peak width determined by $\lambda\propto t'$ as predicted from Eq.\,(5) is reduced too. More importantly, while the decaying of the PC in isolated trimers is monotonic with the temperature [Fig.\,4(b)], Fig.\,4(c) demonstrates that for a weak but nonzero $\Gamma$ (i.e., $T_{K}<\lambda\ll J$, see the curves with $\Gamma/t=1,\,1.3,$ and $1.4$), the PC peak begins with an appreciable rise before it decays, as $T$ increases from zero. We attribute this counterintuitive increase of the current near criticality to the reduction of Kondo correlations by thermal disturbances. On the right side of the PC peak ($t'\gtrsim t'_c$), the dominant mechanism is $\bm{S}_1$ being Kondo screened by environmental electrons while $\bm{S}_2$ and $\bm{S}_3$ forming a singlet \cite{Mitchell2009, Mitchell2013}. A slight increase in temperature, on the order of fractional $T_\mathrm{K}$, can release $\bm{S}_1$ from the Kondo bound state, resulting in an enhancement of the current as evident in the inset of Fig.\,3(d). However, the primary ferromagnetic Kondo physics on the $t'\lesssim t'_c$ side is the total trimer behaving as a free local moment effectively decoupled from the environment \cite{Mitchell2009, Mitchell2013}, which makes the left side of the PC peak insensitive to the temperature [inset of Fig.\,3(d)]. This explains the unusual increase of the critical currents as the temperature slightly elevates from zero. For larger $\Gamma$ such that $\lambda\ll T_{\mathrm{K}}\sim J$ [see the curves with $\Gamma/t=2$ and $3$ in Fig.\,4(c)], the temperature needed to suppress the Kondo screening of $\bm{S}_1$ is already too high as compared with the energy scale $\lambda$ of the current. Therefore, thermal decoherence predominates, leading to monotonic decay of the critical currents. Such kinds of anomalous PC rise with temperature have also been found in other systems. The anomalous current increase in a Wigner crystal ring with an impurity is due to the enhanced tunneling probability \cite{Zvyagin1995}. The momentum reconstruction in a circular Hubbard model yields temperature-enhanced PCs \cite{Patu2022}. Similar phenomena, e.g. the Schottky anomaly, can occur in finite-size systems due to discrete energy levels. The effect we find has a Kondo-related mechanism, thereby distinct from all these known phenomena.

In conclusion, we have studied the persistent current in a magnetic-atom trimer nanoring coupled to a metallic host. The trimer molecule exhibits a reversal of aromaticity driven by QPTs, distinct from the aromaticity reversal recently predicted in cyclocarbons \cite{Roncevic2023}. Upon tuning the hopping amplitude of frustrated trimers, the paratropic current develops a peak with anomalous temperature dependence, indicating a frustrating quantum criticality between the ferromagnetic and antiferromagnetic Kondo regimes. Our work is helpful for understanding the aromaticity of cyclic molecules beyond the H\"{u}ckel's rule, and informative for exploring loop current orders in more complex systems such as kagome and frustrated Kondo lattices.

\begin{acknowledgments}
This work was supported by the National Natural Science Foundation of China (Grants No.\,12474045, No.\,12274466, No.\,12374034, and No.\,11921005), the Innovation Program for Quantum Science and Technology (2021ZD0302403), the Strategic Priority Research Program of Chinese Academy of Sciences (XDB28000000), the Hunan Provincial Science Fund for Distinguished Young Scholars (2023JJ10058), and the High Performance Computing Center of Central South University.
\end{acknowledgments}


\begin{thebibliography}{61}
\bibitem{note}
Since virtual charge excitations have a gap of $U$, the frustrated trimer must be operated in practical situations with a strong but finite $U$ in order to observe the persisitent current. Otherwise, the current vanishes in the $U\to\infty$ limit. We set a proper interaction, $U=16t$, for numerical results in Figs.\,3 and 4.
\end{thebibliography}
\end{document}